\documentclass{aa}

\usepackage{color}
\usepackage{amsmath}
\usepackage{amssymb}
\usepackage{graphicx}                                                                                

\begin{document}

\newcommand{\abs}[1]{\left\lvert #1 \right\rvert}
\newcommand{\A}[1]{\boldsymbol{#1}}
\newcommand{\sgn}{\textrm{sgn}}
\newcommand{\pd}[2]{\frac{\partial #1}{\partial #2}}
\newcommand{\vhspace}[2]{\vspace{#1}\hspace{#2}}
\newcommand{\Div}[1]{\nabla \cdot #1}
\newcommand{\vgrad}[1]{\boldsymbol{v} \cdot \nabla #1}
\newcommand{\HALF}{\frac{1}{2}}
\newcommand{\vol}{\mathcal{V}}
\newcommand\tabcaption{\def\@captype{table}\caption}
\newcommand{\eq}[1]{(\ref{#1})}
\newcommand{\nv}{\hat{\A{n}}}
\newcommand{\DS}{\displaystyle}
\newcommand{\TS}{\textstyle}
\newcommand{\err}[1]{\color{red} #1 \color{black}}
\newcommand{\new}[1]{\color{blue} #1 \color{black}}

\title{Simulating radiative astrophysical flows with the PLUTO code: \\
       A non-equilibrium, multi-species cooling function}
\titlerunning{A non-equilibrium cooling function for the PLUTO code}

\author{O. Te\c{s}ileanu\inst{1} 
   \and A. Mignone\inst{1}
   \and S. Massaglia\inst{1} }
\institute{Dipartimento di Fisica Generale, Universit\`a degli Studi di Torino, via P. Giuria 1, 10125 Torino, Italy}

\date{Received $<date>$ / Accepted $<date>$}

\abstract
{Time-dependent cooling processes are of paramount importance in the evolution of astrophysical
gaseous nebulae and, in particular, when radiative shocks are present. Given the recent 
improvements in resolution of the observational data, simulating these processes in a more realistic 
manner in magnetohydrodynamic (MHD) codes will provide a unique tool for model discrimination.}
{The present work introduces a necessary set of tools that can be used to model radiative astrophysical 
flows in the optically-thin plasma limit. 
We aim to provide reliable and accurate predictions of emission line ratios
and radiative cooling losses in astrophysical simulations of shocked flows.
Moreover, we discuss numerical implementation aspects to ease future improvements 
and implementation in other MHD numerical codes.}
{The most important source of radiative cooling for our plasma conditions comes from the 
collisionally-excited line radiation. 
We evolve a chemical network, including 29 ion species, to compute the ionization balance 
in non-equilibrium conditions.
The numerical methods are implemented in the PLUTO code for astrophysical fluid dynamics and 
particular attention has been devoted to resolve accuracy and efficiency issues 
arising from cooling timescales considerably shorter than the dynamical ones.
}
{After a series of validations and tests, typical astrophysical setups are simulated in 1D and 2D,
employing both the present cooling model and a simplified one. The influence of the cooling model 
on structure morphologies can become important, especially for emission line diagnostic purposes.}
{The tests make us confident that the use of the presented detailed radiative cooling treatment 
will allow more accurate predictions in terms of emission line intensities and shock dynamics
in various astrophysical setups.}


\maketitle

\section{Introduction}

Astrophysical gases emit thermal radiation while undergoing dynamical transformations.
There are cases where the gas is so diluted that the typical timescales for cooling
greatly exceed the dynamical ones but, in many instances, cooling and the related 
ionization/recombination processes for the emitting species become comparable to, 
or faster than, the dynamical evolution of the system and should be considered. 
Classical examples of intensively radiating gases are \ion{H}{ii} regions, planetary nebulae, 
supernova remnants and star forming regions. 
Thus, when studying gas flows in such environments, particular care must be taken to 
treat the interplay between dynamics and radiation in the correct way. 
This is particularly true and crucial whenever radiative shocks are involved. When 
cooling is strong, the ionization fraction of the emitting species is far from that at 
equilibrium, but evolves so rapidly with time that it must be treated in a time-dependent 
fashion. 

Under these conditions, the magnetohydrodynamic (MHD) equations, coupled with the equations 
describing the evolution of the emitting species and the radiative losses, 
must be solved by numerical means. 
The numerical problems posed by radiative cooling are particularly challenging 
whenever the advance time step of the integration is controlled by radiation/ionization 
rather than dynamics. This may happen in few grid points in the computational domain where the radiative 
losses are intense, e.g., right behind a shock front, where the cooling time becomes very small. 
Therefore, it is necessary to devise strategies able to deal with
very different integration timescales: this is one of the points we address.

Time dependent ionization calculations were previously performed for gaseous nebulae
by Marten \& Szczerba (\cite{MS97}) in hydrostatic conditions. Their approach is
similar to our implementation of the ionization state treatment, however the ion species
and implemented physical processes are in part different. Also, radiative cooling in optically-thin plasmas 
was previously investigated, and synthetic cooling functions were designed 
(e.g., Schmutzler \& Tscharnutter \cite{ST93}).
Among the radiative numerical codes employed in astrophysics, one can quote the 
hydrocode YGUAZ\`U (Raga et al. \cite{RA00}), ASTROBEAR 
MHD code (Poludnenko et al. \cite{PA05}, Berger \& LeVeque \cite{BL98}) 
and Virginia Hydrodynamics - 1 (VH-1, Sutherland et al. \cite{SA03}, Blondin \& Lufkin \cite{BL93}). 
The MHD simulation code we use for our astrophysical applications -- PLUTO -- is a freely
distributed application developed and maintained at the Turin University -- Turin Astronomical Observatory 
(Mignone et al. \cite{MA07}). 
A previous numerical analysis about the evolution of radiative shocks in Young Stellar Object (YSO) jets 
(Massaglia et al. \cite{MA05}) was carried out with PLUTO, using a simplified model for 
the radiative cooling losses, which evolved in time only the ionization fraction 
of hydrogen (c.f., Rossi et al. \cite{RA97}). This model will be called from now on
SNEq (Simplified Non-Equilibrium cooling). 

We illustrate a more general treatment of atomic cooling and evolution of 
the ionization fraction of the emitting species, embedded in the PLUTO code as well, 
for use within MHD simulations of astrophysical interest. We will call this new 
cooling function MINEq (Multi-Ion Non-Equilibrium cooling). The main 
advantage of our approach is the full ionization state computation during
the MHD simulation, which allows for better predictions of emission line intensities.

Section \ref{general} contains a general overview of the adopted method and 
implementation of the treatment of radiative losses. Then, in Sect. \ref{cfdesc}, a 
description of the physics of the cooling model can be found, followed in
Sect. \ref{testeq} by the validations and tests in equilibrium conditions.
The numerical implementation and testing are discussed in Sect. \ref{numerical}, 
while in Sect. \ref{testint} we present some typical astrophysical applications.
Technical details on ionization-recombination processes and 
numerical issues are presented in extended form in the Appendix.

%%%%%%%%%%%%%%%%%%%%%%%%%%%%%%%%%%%%%%%%%%%%%%%%%%%%%%%%%%%%%%%%%%%%%%%%%
\section{General overview}\label{general}
%
%
%
%
%
%
%
%%%%%%%%%%%%%%%%%%%%%%%%%%%%%%%%%%%%%%%%%%%%%%%%%%%%%%%%%%%%%%%%%%%%%%%%%

The general characteristics and application ranges of the new cooling function added 
to the PLUTO code are summarized below. The density limits are those typically
encountered in clouds and YSO jets, while the temperature range is limited 
by the highest ionization stage considered (at the high end) and the lack of molecular 
cooling (at the low end):
\begin{equation}
N \in (10^{-2}, 10^5) \ {\mathrm{cm^{-3}}} \,,\quad
T \in (2 \cdot 10^3, 2 \cdot 10^5) \ {\mathrm{K}} \,.
\end{equation}
However, the module is designed to permit later extension in terms 
of applicable parameter range (through adding more ion species, or a
tabulated cooling function for higher temperatures) and physical processes 
taken into consideration.

Flow variables such as density $\rho$, velocity $\vec{v}$, magnetic field $\vec{B}$, and
total energy $E$ are evolved according to the standard MHD equations:
\begin{eqnarray}
 \pd{\rho}{t} + \nabla\cdot\left(\rho\vec{v}\right) & = & 0  \,
 \label{eq:CL1} \\ 
 \pd{(\rho\vec{v})}{t} + \nabla\cdot\left(\rho\vec{v}\vec{v}^T - \vec{B}\vec{B}^T + \tens{I}p_t\right) &=& 0\,
  \label{eq:CL2} \\
 \pd{\vec{B}}{t} - \nabla\times\left(\vec{v}\times\vec{B}\right) &=& 0 \,
 \label{eq:CL3} \\ 
 \pd{E}{t} + \nabla\cdot\left[\left(E + p_t\right)\vec{v} - \left(\vec{v}\cdot\vec{B}\right)\vec{B}\right]
 & = &S_E   \label{eq:CL4}  \,,
\end{eqnarray}
where $S_E$ (described later) is a radiative loss term, and
$p_{\mathrm{t}} \equiv p + |\vec{B}|^2/2$
denotes the total pressure (thermal + magnetic) of the fluid.
We assume an ideal equation of state by which the total energy density becomes 
\begin{equation}
 E = \frac{p}{\Gamma - 1} + \rho\frac{\left|\vec{v}\right|^2}{2} 
                          + \frac{\left|\vec{B}\right|^2}{2} \,,
\end{equation}
with $\Gamma=5/3$ being the specific heats ratio.

The cooling model accounts for the evolution of 
29 ion species, namely: \ion{H}{i}, \ion{H}{ii}, \ion{He}{i} and 
\ion{He}{ii}, and the first five ionization stages of C, N, O, Ne and S. 
Sulphur, although not having an important contribution 
to cooling, is added for diagnostic purposes (line ratios).
The ionization network employed is larger than 
in most other MHD codes.

For each ion, we solve the additional equation
\begin{equation}\label{eq:species}
 \pd{(\rho X_{\kappa,i})}{t} + \nabla\cdot\left(\rho X_{\kappa,i}\vec{v}\right) = \rho S_{\kappa,i} \,
\end{equation}
coupled to the original system of conservation laws (\ref{eq:CL1})--(\ref{eq:CL4}).
In Eq. (\ref{eq:species}) and throughout the following, the first index ($\kappa$)
corresponds to the element, while the second index ($i$) corresponds to the ionization stage.
Specifically, $X_{\kappa,i} \equiv N_{\kappa,i} / N_{\kappa}$ is the
ion number fraction, $N_{\kappa,i}$ is the number density of the $i$-th ion of element $\kappa$,
and $N_{\kappa}$ is the element number density.
We denote the whole set of ions for all possible $\kappa$ and $i$ 
with $\vec{X}\equiv\{X_{\kappa,i}\}$.

The source term $S_{\kappa,i}$ accounts for ionization and recombination and 
will be described in the following.
The total line emission from these species enters in the source term $S_E$ in Eq. (\ref{eq:CL4})
and should give a good approximation of radiative cooling for the above conditions 
(Raga et al. \cite{RM97}). 

The system of Eqs. (\ref{eq:CL1})--(\ref{eq:CL4}) together with (\ref{eq:species}) 
is integrated using the PLUTO code for computational astrophysics (Mignone et al. \cite{MA07}).
We take advantage of operator splitting techniques, where the homogeneous part 
of the equations (i.e., with $S_{\kappa,i}=S_E=0$) is solved separately from the 
source step.
The order of the respective advection and source operators (${\cal H}^{\Delta t^n}$ and
${\cal S}^{\Delta t^n}$) is reversed every step by keeping 
the time step $\Delta t^n=\Delta t^{n+1}$ constant for two consecutive integrations
to guarantee formal second order accuracy.
Thus, if $\vec{U}=\{\rho, \rho\vec{v}, \vec{B}, E, \rho\vec{X}\}$ is the
vector of conserved variables, the solution advances from $t^n$ to $t^n + \Delta t^n$
as
\begin{equation}\label{eq:split1}
 \vec{U}(t^n + \Delta t^n) = {\cal S}^{\Delta t^n}{\cal H}^{\Delta t^n}\, \vec{U}(t^n)\,,
\end{equation}
and from $t^n + \Delta t^n$ to $t^n + 2\Delta t^n$ as
\begin{equation}\label{eq:split2}
 \vec{U}(t^n + 2\Delta t^n) = {\cal H}^{\Delta t^n}{\cal S}^{\Delta t^n} \, \vec{U}(t^n + \Delta t^n)\,.
\end{equation}
A new time step, $\Delta t^{n+2}$, is then computed as shown in Sect. \ref{numerical}.

%%%%%%%%%%%%%%%%%%%%%%%%%%%%%%%%%%%%%%%%%%%%%%%%%%%%%%%%%%%%%%%%%%%%%%%%%
\section{Cooling module description}\label{cfdesc}
%
%
%
%
%
%
%%%%%%%%%%%%%%%%%%%%%%%%%%%%%%%%%%%%%%%%%%%%%%%%%%%%%%%%%%%%%%%%%%%%%%%%%

We will restrict our attention to the source step only and remind
the interested reader of the original paper by Mignone et al. (\cite{MA07}) for 
implementation details on the solution of the homogeneous MHD equations.

During the source step, in virtue of operator splitting, only internal energy 
$p/(\Gamma-1)$ and ion fractions $X_{\kappa,i}$ are affected. Density, velocity, and 
magnetic fields remain constant with the values provided by
the most recent step. 
Thus Eqs. (\ref{eq:CL4}) and (\ref{eq:species}) are treated as a system 
of ordinary differential equations (ODE):
\begin{equation}\label{eq:source_ode}
 \frac{d}{dt}\left(\begin{array}{c}
   p \\   \noalign{\medskip}
   X_{\kappa,i} 
 \end{array}\right) 
 =
 \left(\begin{array}{c}
   (\Gamma - 1) S_E \\  \noalign{\medskip} 
   S_{\kappa,i}  
 \end{array}\right)  \,,
\end{equation}
where $\kappa = \rm{H, He, C,}...$ labels the element and 
$i=\rm{I,II, III,}...$ identifies the ionization stage.
Equations (\ref{eq:source_ode}) must be solved for a time increment $\Delta t^n$
with initial condition provided by the output of the previous 
step (i.e., either an advection or source one).

Pressure $p$ and temperature $T$ are related by the ideal gas equation:
\begin{equation}\label{eq:eos}
  p = Nk_B T \,\qquad {\rm with} \qquad
  N = \frac{\rho}{m_u\mu(\vec{X})} \,,
\end{equation}
where $N$ is the total particle (atoms + electrons) number density, 
$k_B$ is the Boltzmann constant, $m_u$ is the atomic mass unit, 
and $\mu(\vec{X})$ is the mean molecular weight:
\begin{equation}\label{eq:mu}
  \mu ( \vec{X} ) = \frac{\sum_{\kappa} m_{\kappa} B_{\kappa}}{\sum_{\kappa}\sum_i X_{\kappa,i}\gamma_i B_{\kappa}} \,.
\end{equation}
In Eq. (\ref{eq:mu}) $m_{\kappa}$ is the atomic mass (in units of $m_u$) 
of element $\kappa$ and $\gamma_i$ denotes the number of the ionization stage 
in spectroscopic notation for each ion.
$B_{\kappa}$ is the fractional abundance of the element.

%%%%%%%%%%%%%%%%%%%%%%%%%%%%%%%%%%%%%%%%%%%%%%%%%%%%%%%%%%%%%%%%%%%%%%%%%
\subsection{Radiative losses} \label{radloss}
%
%
%
%
%
%%%%%%%%%%%%%%%%%%%%%%%%%%%%%%%%%%%%%%%%%%%%%%%%%%%%%%%%%%%%%%%%%%%%%%%%%

Radiative losses are described by the source term $S_E$ in the energy Eq. (\ref{eq:CL4}):
\begin{equation}
 \label{eq:en_source}
 S_E = -\Big(N_{\rm at}N_{\rm el}\Lambda\left(T,\vec{X}\right) + L_{\rm FF} + L_{\rm I-R}\Big)\,, 
\end{equation}
where $\Lambda(T, \vec{X}) $ is the radiative cooling function due to 
collisionally-excited line radiation, $L_{\rm FF}$  
denotes the free-free (bremsstrahlung) losses from H$^{\mathrm{+}}$ and He$^{\mathrm{+}}$, 
while $L_{\rm I-R}$ accounts for the energy lost during 
ionization/recombination processes.
The number densities $N_{\rm at}$ and $N_{\rm el}$ are, respectively, the total atom and electron number 
densities, readily determined from the mass density and the known chemical composition 
of the plasma (by default supposed solar, but customizable by the user):
\begin{equation}\label{eq:nat}
  N_{\mathrm{at}} = \sum_{\kappa} N_{\kappa}  \,,
\end{equation}
\begin{equation}\label{eq:nel}
  N_{\mathrm{el}}(\vec{X}) = N \sum_{\kappa} \sum_i X_{\kappa,i} ( \gamma_i - 1) B_{\kappa}  \,.
\end{equation}
Note that $N_{\rm at}$ \emph{does not} depend on the ionization state
of the elements and it should not be confused with $N$, the 
total particle (atoms + electrons) number density used in the equation of state
(\ref{eq:eos}).

Emission lines from \ion{Fe}{ii}, \ion{Si}{ii}, and \ion{Mg}{ii} that exist in SNEq 
are added empirically to the energy losses of MINEq 
(without evolving the respective ion species) because of their importance at low temperatures. 

%%%%%%%%%%%%%%%%%%%%%%%%%%%%%%%%%%%%%%%%%%%%%%%%%%%%%%%%%%%%%%%%%%%%%%%%%
\subsubsection{Energy loss by collisionally-excited line radiation}
%
%
%
%
%%%%%%%%%%%%%%%%%%%%%%%%%%%%%%%%%%%%%%%%%%%%%%%%%%%%%%%%%%%%%%%%%%%%%%%%%

The main contribution to radiative cooling comes from collisional excitation of 
low-lying energy levels of common ions, such as O and N. 
In spite of their low abundances, these ions make a 
significant contribution because they have energy levels with excitation 
potentials of the order of $kT$. 
The total radiative cooling function $\Lambda(T,\vec{X})$ used in the energy source 
term (Eq. (\ref{eq:en_source})) is:
\begin{equation}\label{eq:pp}
  \Lambda(T,\vec{X}) = \sum_{\kappa} \sum_i X_{\kappa,i} {\cal L}_{\kappa,i}(N_{\mathrm{el}}, T) B_{\kappa}  \,,
\end{equation}
where the sums are extended to all ion species and $B_{\kappa}$ is the fractional 
abundance of the element $\kappa$. 

Individual contributions to the different ${\cal L}$'s \footnote{in this section only, 
$\kappa$ and $i$ will be omitted unless necessary to avoid cluttered notation} are 
given by 
\begin{equation}\label{eq:Ljdefine}
 {\cal L} = \sum_j \hat{N}_j \sum_{l<j}A_{jl}h\nu_{jl}  \,,
\end{equation}
where $\hat{N}_j$ is the population of the $j$-th excitation level; $A_{jl}$ are the Einstein 
A coefficients; and $\nu_{jl}$ the emission line frequency
for a transition between levels $j$ and $l$.
We consider a 5-level atom model to compute the line radiation 
(Osterbrock \& Ferland \cite{OF05}) by solving for the equilibrium 
populations in each of the excitation levels $j = 1\dots 5$:
\begin{equation}
  \sum_{l\neq j} \hat{N}_lN_{\mathrm{el}}q_{lj} + \sum_{l>j} \hat{N}_lA_{lj} = 
  \sum_{l\neq j} \hat{N}_jN_{\mathrm{el}}q_{jl} + \sum_{l<j} \hat{N}_jA_{jl} \,,
\end{equation}
which, together with the normalization condition for the total number density of the ion,
$\sum_j \hat{N}_j = N_{\kappa,i}$, can be solved for the relative 
$\hat{N}_j$ populations in each level.

The 5-level atom model provides the great majority of the emission lines 
for the considered range of temperatures and thus gives a reliable estimation
of the total line cooling.

For most of the ion species, the emission coefficients were taken from Pradhan \& Zhang 
(\cite{PZ99}). The data for hydrogen was taken from Giovanardi et al. 
(\cite{GA87}) and their fit formula. The \ion{C}{ii} data comes from Blum \& Pradhan (\cite{BP92}), 
while for \ion{N}{ii} and \ion{N}{iii} Chebyshev polynomial fits from Stafford et al. (\cite{SA94}) 
were used.

%%%%%%%%%%%%%%%%%%%%%%%%%%%%%%%%%%%%%%%%%%%%%%%%%%%%%%%%%%%%%%%%%%%%%%%%%
\subsubsection{Free-free radiation}
%
%
%
%
%
%%%%%%%%%%%%%%%%%%%%%%%%%%%%%%%%%%%%%%%%%%%%%%%%%%%%%%%%%%%%%%%%%%%%%%%%%

A minor contributor to the cooling rate at moderate temperatures is the bremsstrahlung 
(free-free) radiation, having a continuous spectrum. The rate of cooling in this process 
by ions of charge $Z$, integrated over frequency, is approximately 
(Osterbrock \& Ferland \cite{OF05})
\begin{equation}\label{eq:Lffdefine}
  L_{\rm FF} = 1.42 \times 10^{-27} Z^2 T^{1/2} N_{\rm el} N_+ \,,
\end{equation}
in $\mathrm{ergs\,cm^{-3}\,s^{-1}}$. Because of its abundance, H$^+$ dominates the 
free-free cooling, and He$^+$ can be included along with it since both have the same charge: 
$N_+ \equiv N_{\rm HII} + N_{\rm HeII}$.

%%%%%%%%%%%%%%%%%%%%%%%%%%%%%%%%%%%%%%%%%%%%%%%%%%%%%%%%%%%%%%%%%%%%%%%%%
\subsubsection{Ionization-recombination losses}
%
%
%
%
%
%%%%%%%%%%%%%%%%%%%%%%%%%%%%%%%%%%%%%%%%%%%%%%%%%%%%%%%%%%%%%%%%%%%%%%%%%

Thermal energy is absorbed by the atom to pass to the next ionization stage.
During the recombination, a free electron is captured and a part of its
thermal energy is lost. The ionization/recombination losses are treated similarly in MINEq and SNEq, 
following the method described in Rossi et al. (\cite{RA97}):
\begin{eqnarray}
  L_{\rm I-R}  & = & L_{\rm I} + L_{\rm R}                                     \nonumber      \\
  L_{\rm I}    & = & 1.27 \cdot 10^{-23} \sqrt{T} N_{HI} e^{-\frac{157890}{T}}  N_{\rm el}    \\
  L_{\rm R}    & = & 2.39 \cdot 10^{-27} \sqrt{T} N_{HII} N_{\rm el}       \nonumber 
\end{eqnarray}
expressed in ${\rm ergs\,cm^{-3}\,s^{-1}}$.

The complementary effects of these processes on the plasma (the creation/destruction of a free particle) 
are accounted through the mean molecular weight, which varies together with
the total particle number density.

%%%%%%%%%%%%%%%%%%%%%%%%%%%%%%%%%%%%%%%%%%%%%%%%%%%%%%%%%%%%%%%%%%%%%%%%%%
\subsection{Ionization network}\label{eqbal}
%
%
%
%
%%%%%%%%%%%%%%%%%%%%%%%%%%%%%%%%%%%%%%%%%%%%%%%%%%%%%%%%%%%%%%%%%%%%%%%%%%

Our ionization network can be written in terms of the source-term $S_{\kappa,i}$ mentioned 
above (Eq. (\ref{eq:source_ode})):
\begin{equation}\label{eq:ion_bal}
S_{\kappa,i}  = N_{\mathrm{el}} \Big[ X_{\kappa,i+1} \alpha_{\kappa,i+1} - X_{\kappa,i} \left( \zeta_{\kappa,i}
                + \alpha_{\kappa,i} \right) 
    + X_{\kappa,i-1} \zeta_{\kappa,i-1} \Big]    \,,
\end{equation}
where $\zeta_{\kappa,i}$ and $\alpha_{\kappa,i}$ are the ionization and 
recombination coefficients of the $i$-th ion specie of the element $\kappa$, defined as follows:
\begin{equation}\label{eq:coeffI}
\zeta_{\kappa,i} = \zeta^{\rm coll}_{\kappa,i} (T) + \frac{N_{\rm HII}}{N_{\rm el}}  
                   \zeta^{\rm HII}_{\kappa,i} (T)   \,,
\end{equation}
\begin{equation}\label{eq:coeffR}
\alpha_{\kappa,i} = \alpha^{\rm el}_{\kappa,i}(T) + \frac{N_{\rm HI}}{N_{\rm el}}  \alpha^{\rm HI}_{\kappa,i} (T) 
    + \frac{N_{\rm HeI}}{N_{\rm el}} \alpha^{\rm HeI}_{\kappa,i} (T)     \,,
\end{equation}
where $N_{\rm HII} \equiv N_{\rm H} X_{\rm HII} = N_{\rm at}B_{\rm H}X_{\rm HII}$ is the number density of protons, 
$N_{\rm HI}$ and $N_{\rm HeI}$ are the number densities of neutral hydrogen and 
helium, respectively. The $\alpha_{\kappa,i}$ and $\zeta_{\kappa,i}$ coefficients 
are the transition rates corresponding to the reaction mechanisms defined in the Appendix A 
(note that $\alpha^{\rm el}_{\kappa,i}$ is the total electron-ion recombination coefficient, 
that is dielectronic \emph{plus} radiative, $\alpha^{\rm el} = \alpha^{\rm DR} + \alpha^{\rm RR}$).

Since we only consider part of the ions from each element (up to the fourth level of ionization, except for H and He), 
the ionization rate for the highest state will be set to zero. This will produce saturation of the ion
population in this state at very high temperatures, and limit the applicability of the cooling function. 
The temperature range can however be extended by adding further ionization stages for the elements. 

For efficiency purposes, the ionization and recombination coefficients on the
right-hand side of Eqs. (\ref{eq:coeffI}) and (\ref{eq:coeffR}) are sampled at discrete
values of temperature at the beginning of integration and used as lookup tables.

%%%%%%%%%%%%%%%%%%%%%%%%%%%%%%%%%%%%%%%%%%%%%%%%%%%%%%%%%%%%%%%%%%%%%%%%%%%%%%%%%%%%%%%%%%
\section{Comparison with equilibrium models}\label{testeq}
%
%
%
%
%
%
%
%%%%%%%%%%%%%%%%%%%%%%%%%%%%%%%%%%%%%%%%%%%%%%%%%%%%%%%%%%%%%%%%%%%%%%%%%%%%%%%%%%%%%%%%%%

We perform theoretical line ratios tests to verify the collision strengths in the 
radiative losses. Also, the total cooling function for an equilibrium ionization balance function 
of temperature (the effective cooling curve) was tested and found to be consistent with results 
obtained with more complex models.

%%%%%%%%%%%%%%%%%%%%%%%%%%%%%%%%%%%%%%%%%%%%%%%%%%%%%%%%%%%%%%%%%%%%%%%%%
\subsection{Equilibrium ionization balance}
%
%
%
%
%%%%%%%%%%%%%%%%%%%%%%%%%%%%%%%%%%%%%%%%%%%%%%%%%%%%%%%%%%%%%%%%%%%%%%%%%

The equilibrium ionization balance may be used as an initial condition for  
numerical simulations, and also serves for testing the ionization/recombination
coefficients employed in the ionization network.

The ionization balance for each element at equilibrium is computed by setting 
$dX_{\kappa,i} / dt = 0$ (for all ions) in Eq. (\ref{eq:source_ode}). 
The equation for the highest ionization stage is replaced by the normalization 
condition,
\begin{equation}\label{eq:norm}
\sum_{i=1}^{K} X_{\kappa,i} = 1   \,,
\end{equation}
where $K$ is the highest ionization state taken in consideration for the element $\kappa$.
Thus, for each element, we solve the following system of equations:
\begin{equation}\label{eq:ion_eq_system}
 X_{\kappa,i+1} \alpha_{\kappa,i+1} - X_{\kappa,i} \left( \zeta_{\kappa,i} + \alpha_{\kappa,i} \right) 
 +  X_{\kappa,i-1} \zeta_{\kappa,i-1} = 0 \,, 
\end{equation}
with $i=1,\cdots,K-1$ complemented by Eq. (\ref{eq:norm}).
Despite its aspect, the previous system of equations is not linear since the 
$\zeta$ and $\alpha$ coefficients depend on the concentrations themselves (see Eqs. (\ref{eq:coeffI}) and
(\ref{eq:coeffR})), so an iterative procedure must be employed to converge to the correct solution.

In the particular cases of hydrogen and helium, because of the charge-transfer reactions they are involved in, 
an exact treatment would also force $\zeta$ and $\alpha$ to depend on the number 
densities of all other ions that take part in these processes. Considering the very limited influence of 
these reactions on the hydrogen and helium ionization balance, we chose to neglect such influences.

\begin{figure} [!ht]
  \resizebox{\hsize}{!}{\includegraphics{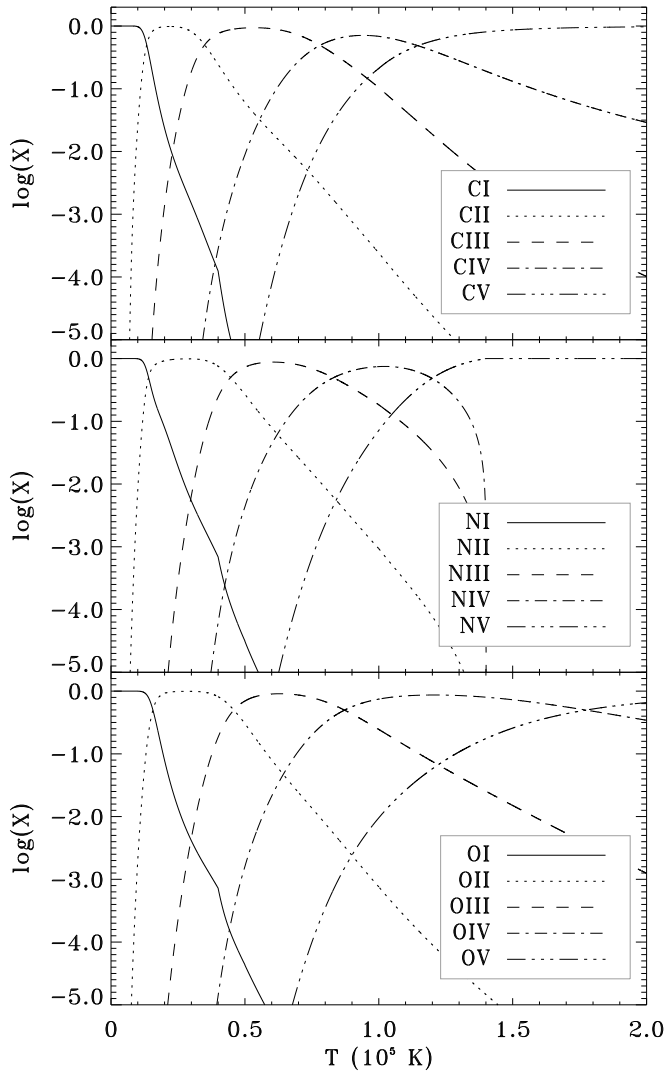}}
  \caption{Ionization fractions at equilibrium for the five ionization levels considered 
           for C (top panel), N (middle panel), and O (bottom panel).}
  \label{fig:eq_bal}
\end{figure}

Given an initial guess on the ionization state of the plasma, the systems of equations for equilibrium are solved,
providing new values of $N_{\mathrm{el}}$, $N_{\mathrm{HI}}$, and $N_{\mathrm{HeI}}$.
The process is repeated until the differences between the old and the new solutions are below a 
certain threshold. The convergence is rapidly achieved, 
generally less than five iterations are needed for a $10^{-4}-10^{-3}$ relative threshold. 
This is acceptable, considering that this equilibrium computation is typically done on the 
whole computation grid only once, in the beginning of the simulation.
In Fig.~\ref{fig:eq_bal}, we show the equilibrium ionization balance as a function 
of temperature for three selected elements.
Our results favourably compare to those obtained by previous investigators--such as 
Sutherland \& Dopita (\cite{SD93})--with the ionization fractions 
being within $5-10\%$ at the same temperature. 

%%%%%%%%%%%%%%%%%%%%%%%%%%%%%%%%%%%%%%%%%%%%%%%%%%%%%%%%%%%%
\subsection{Line ratios tests}
%
%
%
%
%
%%%%%%%%%%%%%%%%%%%%%%%%%%%%%%%%%%%%%%%%%%%%%%%%%%%%%%%%%%%%

These tests are useful to verify the emission lines data 
and the level population computation routine (in our 5-level atom model). 
An example is presented here.

A popular way of estimating the temperatures in gaseous nebulae is to use the 
ratio of spectral line intensities, such as the lines of \ion{O}{iii}:
\begin{equation}
  \frac{\epsilon(\lambda 5007)+\epsilon(\lambda 4959)}{\epsilon(\lambda 4363)} = 
  \frac{\epsilon(^1D_2 \rightarrow {}^3P_2) + \epsilon(^1D_2 \rightarrow {}^3P_1)}{\epsilon(^1S_0 \rightarrow {}^1D_2)}
 \,.
\end{equation}
Inserting numerical values of the collision strengths and transition probabilities 
(Osterbrock \& Ferland \cite{OF05}), the ratio becomes:
\begin{equation}
\label{eq:analytic_OIII}
  R = \frac{\epsilon(\lambda 5007)+\epsilon(\lambda 4959)}{\epsilon(\lambda 4363)} = 
  \frac{8.32 \exp \left( \cfrac{3.29\times 10^4}{T} \right)}{1 + 4.5 \times 10^{-4}\cfrac{N_{\rm el}}{T^{1/2}}} \,,
\end{equation}
for temperatures around 10\,000K.

Line ratios computed with the previous formula and the results of the 5-level atom model 
were compared, the differences of less than $\approx 6\%$ being due to the fact that our 
code uses temperature-dependent collision strengths, while in the formula above they are 
assumed constant.

%%%%%%%%%%%%%%%%%%%%%%%%%%%%%%%%%%%%%%%%%%%%%%%%%%%%%%%%%%%%%
\subsection{Effective cooling}\label{effcool}
%
%
%
%
%
%%%%%%%%%%%%%%%%%%%%%%%%%%%%%%%%%%%%%%%%%%%%%%%%%%%%%%%%%%%%%

\begin{figure} [!ht]
  \begin{center}
      \resizebox{\hsize}{!}{\includegraphics{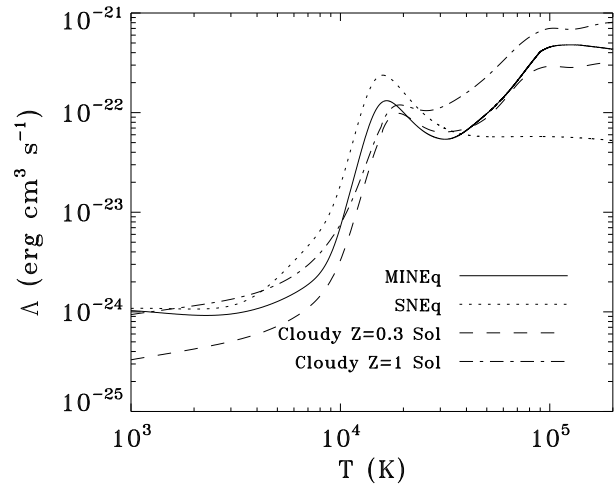}}
    \caption{Effective cooling curves in the temperature range from $10^3$ to $2\cdot 10^5$K, 
             comparison between the results obtained with MINEq, SNEq, and Cloudy.}
 \label{fig:total_cool}
 \end{center}
\end{figure}

In Fig.~\ref{fig:total_cool}, we plot the effective cooling function 
(in $\mathrm{erg\,cm^3 \,s^{-1}}$) using the ionization fractions computed at
equilibrium.
For the sake of comparison, we also show the results obtained with the SNEq model described in 
Rossi et al. (\cite{RA97}) and the Cloudy atomic code, which has a large chemical network 
(see Ferland et al. \cite{FA98}), for similar plasma conditions. 
Solar abundances have been assumed for all cooling functions, except for the Cloudy $Z=0.3$ case
(where the metallicity is only $0.3$ times the solar one).
The SNEq model consists of the emission of 17 most important lines, plus the two-photon continuum
and the radiative losses from ionization/recombination processes. In this model, however, only the 
ionization of H is evolved with the integration, the rest of the ions abundances being fixed or 
locked by charge-transfer processes to the ionization state of H.

The results show a good agreement between the newly-developed cooling 
model (MINEq) and the computations carried out with the Cloudy code. 
Chemical composition (more extended in Cloudy that is an atomic code) and
physical processes considered account for the differences.
The MINEq effective cooling, considering only few metals, generally lies in between the results obtained 
with the Cloudy code for $Z = 0.3 Z_{\sun}$ and $Z = Z_{\sun}$. 
The faster increase in the peak at 17\,000K due to hydrogen Ly$\alpha$ presented 
by MINEq is due to the different sources of the ionization/recombination and emission
coefficients (collision strengths).

It can be inferred that while MINEq accounts with good accuracy for the cooling losses 
up to $2\cdot 10^5$ K, SNEq cannot follow them above $3\cdot 10^4$ K because it lacks 
higher ionization stages for the atoms. 
Furthermore, the effective cooling obtained with the MINEq model closely reproduces 
the early work of Dalgarno \& McCray (\cite{DM72}) in the temperature range considered.

%%%%%%%%%%%%%%%%%%%%%%%%%%%%%%%%%%%%%%%%%%%%%%%%%%%%%%%%%%%%%%%%%%%%%%%%%
\section{Numerical implementation}\label{numerical}
%
%
%
%
%
%
%
%%%%%%%%%%%%%%%%%%%%%%%%%%%%%%%%%%%%%%%%%%%%%%%%%%%%%%%%%%%%%%%%%%%%%%%%%

In the source step, we advance the system of ordinary differential 
equations (ODE) given by Eq. (\ref{eq:source_ode}) in each computational 
zone.
For ease of notations, we rewrite the system as 
\begin{equation}\label{eq:ode}
 \frac{d\vec{y}}{dt} = \vec{f}(\vec{y}) \,,
\end{equation}
where $\vec{y} \equiv \{p, X_{\kappa,i}\}$ and $\vec{f} \equiv \{(\Gamma-1)S_E, S_{\kappa,i}\}$
are, respectively, the vector of unknowns and right-hand sides for all possible values 
of $\kappa$ and $i$ in a given computational cell.
According to the notations introduced in Eq. (\ref{eq:split1})
and (\ref{eq:split2}), we write the formal solution to (\ref{eq:ode}) 
for a time increment $\Delta t^n$ as $\vec{y}^{*} = {\cal S}^{\Delta t^n}\vec{y}^0$,
where the initial condition $\vec{y}^0$ is given by the output of
the previous step.

%%%%%%%%%%%%%%%%%%%%%%%%%%%%%%%%%%%%%
\subsection{Integration Strategy}
%
%
%
%%%%%%%%%%%%%%%%%%%%%%%%%%%%%%%%%%%%%

Accurate numerical integrations of Eq. (\ref{eq:ode}) should be carried out 
consistently with the different timescales that may concurrently co-exist,
according to the initial density, temperature and chemical concentrations. 
In addition, the system evolution dictated by the local ionization, 
recombination, and cooling rates may proceed considerably faster than
the typical time scale imposed during the advection step.
Under some circumstances, this contrast may lead to a stiff system 
of ODE.
A common occurrence takes place, for instance, when a strong shock propagates 
in a cold neutral medium: as the front advances from one computational 
cell to the next, the ion populations will try to re-adjust to the sudden 
increase in temperature at a rate given by the ionization coefficients.
At high temperatures ($T \gtrsim 2\times 10^4$), this process may proceed  
more and more rapidly.

\begin{figure} [!ht]
  \begin{center}
      \resizebox{6cm}{!}{\includegraphics{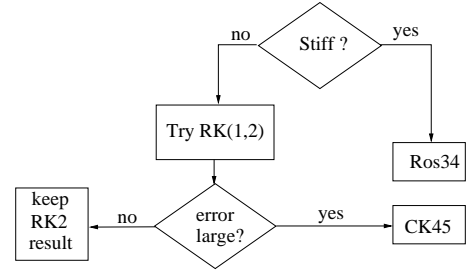}}
    \caption{Flowchart of the dynamically switching integration algorithm 
             for cooling: Runge-Kutta 12, Rosenbrock 34, and Cash-Karp 45.
             Stiffness is detected according to Eq. (\ref{eq:stiff_condition}).}
 \label{fig:int_diagram}
 \end{center}
\end{figure}

Nevertheless, these kinds of events are extremely localized in space since most 
regions ahead of and far behind the shock wave are either close to equilibrium 
or evolve on much slower recombination scales. 
This suggests some form of selective integration by which regions of the flow 
undergoing very rapid changes should be promptly detected and treated accordingly.
We achieve this by 1) detecting potential ``stiffness" due to large  
ionization and recombination coefficients given by Eqs. 
(\ref{eq:coeffI})-(\ref{eq:coeffR})
and 2) monitoring, in each computational cell, the accuracy through an estimate 
of the local truncation error.
We now describe in detail the numerical implementation of a dynamically-adaptive
integration strategy, also shown in Fig.\,\ref{fig:int_diagram}.

At the beginning of integration, we tag a computational cell as ``non-stiff" if 
the integration time step satisfies
\begin{equation}\label{eq:stiff_condition}
  \Delta t^n < \frac{1}{\DS N_{\rm el}
             \max_{\kappa,i}\left(\left|\zeta_{\kappa,i} + \alpha_{\kappa,i}\right|\right)} \,
\end{equation}
where $N_{\rm el}$ is computed in the considered cell.
If the previous condition holds\footnote{
This is, in fact, half the stability limit for the $1^{\rm st}$ order 
explicit Euler method.}, 
we solve Eq. (\ref{eq:ode}) using an explicit method 
with adaptive stepsize control.
Embedded Runge-Kutta (RK) pairs simultaneously giving solutions of order
$m$ and $m-1$ are preferred, since they provide an efficient
error estimate.
The most simple (2,1) pair ($m=2$), for example, may be obtained using 
a simple combination of two right-hand side evaluations
yielding, respectively, $1^{\rm st}$- and $2^{\rm nd}$-order 
accurate solutions $\vec{y}_1$ and $\vec{y}_2$:
\begin{eqnarray}
  \vec{y}^1 &=& \vec{y}^0 + \Delta t^n\vec{f}\left(\vec{y}^0\right) \,, \label{eq:euler} \\
  \vec{y}^2 &=& \vec{y}^0 + \Delta t^n\vec{f}\left(\vec{y}^0
                + \frac{\Delta t^n}{2}\vec{f}^0\right) \label{eq:rk2}\,.
\end{eqnarray}

The difference between the two solutions, $\vec{y}^1$ and $\vec{y}^2$, 
estimates the truncation error of the lower order method,
$O(\Delta t^2)$ for $m=2$. 
The solution given by $\vec{y}^2$ (RK2) is accepted only
if the error falls below some predefined tolerance $\epsilon_{\rm tol}$
(typically $10^{-5}$):
\begin{equation}\label{eq:error}
  \max\left[\left|\frac{p^1}{p^2} - 1\right|, \;
            \max_{\kappa,i}\left(\left|X^1_{\kappa,i} - X^2_{\kappa,i}\right|\right)\right] 
 < \epsilon_{\rm tol}  \,,
\end{equation}
where $(\kappa,i)$ extends to all ion species. A more accurate Runge-Kutta
$(3,2)$ pair may be used instead.
The condition (\ref{eq:error}) is usually satisfied in regions close
to equilibrium ionization balance.
If Eq. (\ref{eq:error}) is not fulfilled we switch to an explicit 
Runge-Kutta method of order $5$ with 
an embedded $4^{\rm th}$ order solution with coefficients
given by Cash-Karp, see Press et al. (\cite{NR}), from now on CK45.
The adaptive strategy provides a $5^{\rm th}$-order accurate solution and
allows us to split (if required) the full time step $\Delta t^n$ into a number 
of smaller sub-steps until the condition
(\ref{eq:error}) is fulfilled in each one of them.

On the other hand, if Eq. (\ref{eq:stiff_condition}) is not met,
explicit time marching may potentially become unstable.
In such situations, integration is carried using a 
$4^{\rm th}$ order semi-implicit Rosenbrock method with a 
$3^{\rm rd}$ order embedded error estimation (Ros34 henceforth).
Rosenbrock schemes can be considered linearly implicit generalizations of 
Runge-Kutta methods, the prototype of which is the semi-implicit
backward Euler method,
\begin{equation}\label{eq:semi-impl}
 \left(\tens{I} - \Delta t^n\tens{J}\right)\cdot
 \left(\vec{y}^{1} - \vec{y}^0\right) = \Delta t^n \vec{f}\left(\vec{y}^0\right) \,.
\end{equation}
These methods retain stability for large time steps at the additional 
cost of computing the Jacobian matrix 
$\tens{J}=\partial\vec{f}/\partial\vec{y}$ of the system and performing
matrix inversions by LU decomposition.
Both features are notoriously time consuming for moderately large 
systems of equations, such as the one we deal with here.
In Appendix (\ref{append_impl}) we show how the Jacobian can be computed using 
combined analytical and numerical differentiation.
The full integration strategy is schematically illustrated in Fig.\,\ref{fig:int_diagram}.

Alternatively, we also found satisfactory results by dividing the whole
step $\Delta t^n$ into smaller ones and by sub-cycling with the explicit 
CK45 scheme.
The sub-time stepping strategy has proved to handle the moderate
stiffness arising at high ($T > 10^5$) temperatures, when the reaction 
rates become large. This makes, in our experience and for the tests
presented in this work, the explicit scheme 
competitive with the semi-implicit method, inasmuch stiffness is 
spatially confined to a small fraction of the computational domain.

Once acceptable solutions $\vec{y}^*$ have been produced in every 
computational zone, we estimate the next time step according to the CFL 
stability restriction and the maximum fractional change produced during the 
radiation step:
\begin{equation}\label{eq:dt_n2}
 \Delta t^{n+2} = \min\left(\Delta t_{\rm adv}, \epsilon_{\max} \Delta t_{\rm rad}\right) \,,
\end{equation}
where, consistently with Eqs. (\ref{eq:split1}) and (\ref{eq:split2}), the minimum 
is taken over two consecutive time steps and 
$\Delta t_{\rm adv}$ is computed from the CFL condition. 
The quantity $\epsilon_{\max}$ specifies the maximum fractional change 
tolerance (typically $0.01 \lesssim \epsilon_{\max} \lesssim 0.1 $) allowed during the
source step.
The radiative time step $\Delta t_{\rm rad}$ is computed as
\begin{equation}
 \Delta t_{\rm rad} = \frac{\Delta t^n}
    {\DS \max_{\rm xyz}\left[\left|\frac{p^0}{p^*} - 1\right|, 
     \max_{\kappa,i}\left(\left|X^*_{\kappa, i} - X^0_{\kappa,i}\right|\right)\right]} \,,
\end{equation} 
Note that small values of $\epsilon_{\rm max}$ will result in a better 
coupling between the advection and source steps, at the cost of reducing the 
overall time step.  

In terms of right-hand side evaluations, the computational overheads 
introduced by the selected algorithms (RK2:CK45:Ros34) are
in the ratio $2:6:3$ in each cell for a given time step. 
The semi-implicit method Ros34 requires,
however, $2$ additional right-hand side calls to form the Jacobian 
(see Appendix \ref{append_impl}) and the inversion of a matrix by LU decomposition.
This makes Ros34 the most computational expensive scheme of integration.

Nevertheless, extensive testing confirms that only a very small fraction 
of computational zones (usually $\lesssim 1\%$)
does actually require this special, but nonetheless crucial, treatment.
The remaining vast majority of cells can be accurately evolved using a second 
order method.
On the other hand, unconditional use of the CK45 or Ros34 algorithms  
throughout the whole grid leads to a noticeable loss of 
computational efficiency with no gain on the overall accuracy.

%%%%%%%%%%%%%%%%%%%%%%%%%%%%%%%%%%%%%%%%%%%%%%%%%%%%%%%%%%%%%
\subsection{Accuracy comparison}
%
%
%
%
%
%%%%%%%%%%%%%%%%%%%%%%%%%%%%%%%%%%%%%%%%%%%%%%%%%%%%%%%%%%%%%

In order to test the accuracy and efficiency of the selected time marching schemes
adopted during the source step, we consider the evolution of a single parcel 
of gas departing from initial conditions far from equilibrium. 
This situation is typically encountered, for example, when a 
strong shock propagates into a cold medium. Neutral atoms
crossing the front will suddenly feel the sharp increase in temperature
and will try to readjust to the new conditions. The ionization timescale 
will be, most likely, much shorter than the typical advection 
scale. One is interested in performing the simulations at
the timestep given by Eq. (\ref{eq:dt_n2}), but this can violate the
condition expressed by Eq. (\ref{eq:stiff_condition}).

We consider two cases in which a single computational zone
is being evolved in time. 
Initial parameters have been found by running a full shock simulation 
like the one presented in Sect. \ref{prop_shocks}, and
selecting the computational zone showing the most extreme stiff conditions,
according to Eq. (\ref{eq:stiff_condition}).
We perform a number of time integrations at constant 
step size using the Euler, RK2, CK45, and Ros34 algorithms
previously described.
Errors are computed with respect to a reference solution obtained with the CK45 integrator with 
a very stringent tolerance ($10^{-8}$) and a small timestep 
($\sim 10^{-4}$ of the cooling timescale):
\begin{equation}
  \epsilon = \frac{\sum_{\kappa,i} \left| X_{\kappa,i} - X^{\mathrm{ref}}_{\kappa,i} \right| }{\sum_{\kappa,i} X^{\mathrm{ref}}_{\kappa,i}} \,.
\end{equation}
The errors in temperature are lower in all cases because the equations of
the chemical network can be, and usually are, more prone to very rapid 
variations (stiffness).

\begin{figure} [!ht]
 \resizebox{\hsize}{!}{\includegraphics{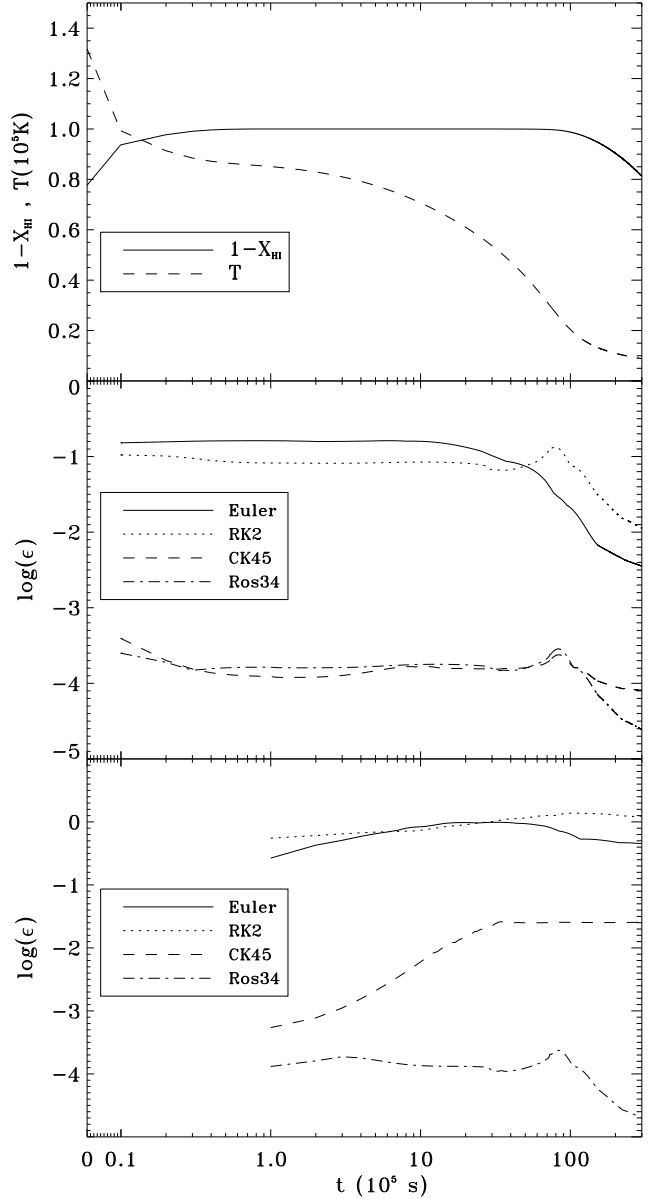}}
 \caption{Top panel: Temperature and ionization fraction of hydrogen evolution; 
          Middle panel: relative errors in ionization fractions, for a 
           fixed timestep $\Delta t = 5 \tau$; 
	  Bottom panel: same as Middle, for a timestep $\Delta t = 50 \tau$.
	  The error plots begin at a time $\Delta t$ from the beginning of the simulation. 
	  Linear time axis up to $10^4s$, logarithmic after.}
 \label{fig:case1}
\end{figure}

In the first case (top panel of Fig.~\ref{fig:case1}), the initial temperature 
is set to $T = 132\,000\,\rm{K}$, the initial neutral hydrogen fraction is 
$22\%$ and the rest of the elements are in the highest ionization stage. 
Under these conditions, the ionization/recombination timescale is 
$\tau \approx 2\cdot 10^3 \,{\rm s}$, typically much smaller
than the scale on which hydrodynamical variables are transported, $\Delta t$.

At a fixed timestep $\Delta t = 5\tau$
(see middle panel in Fig.~\ref{fig:case1}) RK2 is less accurate than Euler, 
this being a typical sign of stiffness (Ekeland et al. \cite{EOO98}). 
The integrator Ros34 yields the best accuracy, immediately followed by CK45.
As the time step is further increased ($\Delta t = 50\tau$, see 
bottom panel in Fig.~\ref{fig:case1}), CK45 progressively loses accuracy, whereas
Ros34 keeps the smallest errors. In this case, the use of a
semi-implicit method clearly reveal its advantages.

\begin{figure} [!ht]
 \resizebox{\hsize}{!}{\includegraphics{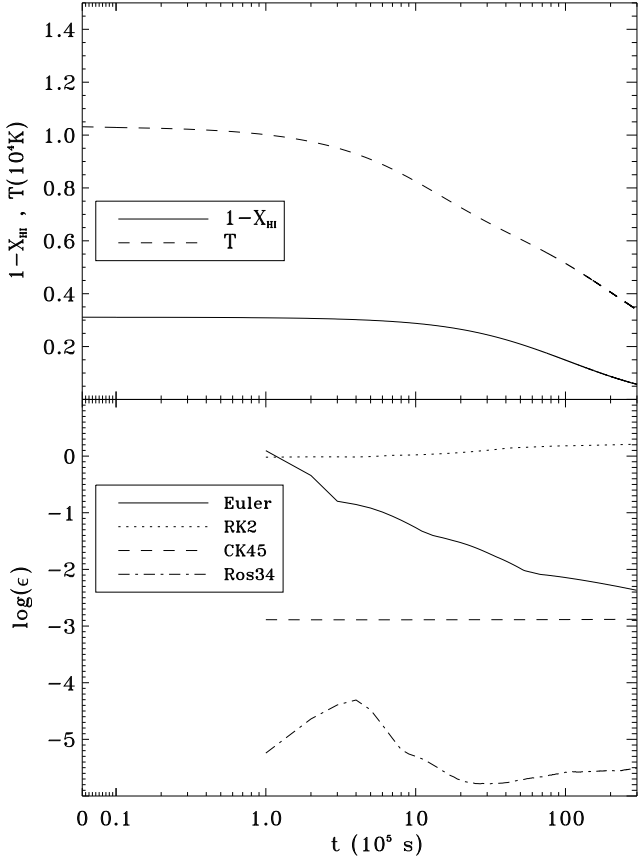}}
 \caption{Top panel: Temperature and ionization fraction of hydrogen evolution; 
          Bottom panel: relative errors in ionization fractions, for a 
           fixed timestep $\Delta t = 100 \tau$. 
	  Linear time axis up to $10^4s$, logarithmic after.}
 \label{fig:case2}
\end{figure}

In the second test, we consider a fully-ionized gas (except for hydrogen,
$X_{\rm HI} \approx 69 \%$) at low temperature $T = 10^4 \,{\rm K}$.
For this choice of parameters, the recombination timescale is 
even smaller than before, $\tau \approx 10^3 \, {\rm s}$.
Figure~\ref{fig:case2} shows the errors computed with selected integration
algorithms when $\Delta t = 100\tau$. 
The resulting accuracies confirm the trend found for the previous case:
low-order, non-adaptive time marching schemes are \emph{not} suitable 
in conditions far from equilibrium. 
On the contrary, Ros34 being an adaptive semi-implicit scheme, 
does not suffer from this loss in accuracy and turns out to be
the best integration method. 
Explicit adaptive algorithms such as CK45, still exhibits somewhat 
better results than the lower order methods. 
It should also be mentioned that the accuracy of explicit
schemes may be further improved if step sub-cycling is used.

We conclude that, for the slow varying regions of the MHD simulation, RK2 (or RK3) can be a 
good choice, while for very fast varying regions a higher-order integrator with time
step adaptivity (as CK45) or even an implicit one (Ros34) become necessary. 
Also, large temperatures are not a necessary condition for the system of equations to become stiff, 
since this can also happen at relatively low temperatures when the ionization/recombination rates are high.

%%%%%%%%%%%%%%%%%%%%%%%%%%%%%%%%%%%%%%%%%%%%%%%%%%%%%%%%%%%%%
\section{Astrophysical applications}\label{testint}
%
%
%
%
%
%%%%%%%%%%%%%%%%%%%%%%%%%%%%%%%%%%%%%%%%%%%%%%%%%%%%%%%%%%%%%

We now apply the newly-developed cooling function to problems
of astrophysical interest.
First, we consider a single, one-dimensional radiative shock 
propagating in a stratified medium with decreasing density.
Then, an example of application of the first setup for the computation
of emission line ratios is shown.
Finally, a study of the dynamical evolution of a jet with varying 
ejection velocity in two-dimensional axial symmetry is presented.

%%%%%%%%%%%%%%%%%%%%%%%%%%%%%%%%%%%%%%%%%%%%%%%%%%%%%%%%%%%%%
\subsection{Propagating shocks}
\label{prop_shocks}
%
%
%
%
%
%%%%%%%%%%%%%%%%%%%%%%%%%%%%%%%%%%%%%%%%%%%%%%%%%%%%%%%%%%%%%

It is interesting to see the difference radiative losses make in the dynamical evolution of a 
propagating shock. A first series of tests were made in 1D, with an initial perturbation in 
pressure, density, and velocity that propagates in a stratified medium of $T_{\rm pre} = 1\,000 K$ and becomes a shock. 
The pre-shock density in the external medium is
\begin{equation}
  \rho_0(x) = \rho_0 \frac{x_0^2}{x_0^2 + x^2} \,,
\end{equation}
where $x$ is the spatial coordinate and the departure density $\rho_0$ corresponds to a particle number
density $N_0 = 10^5 {\rm cm^{-3}}$. This density distribution should approximate well the 
density in an expanding jet. The initial perturbation is set in such a way that only one
shock forms instead of the usual pair of forward/reverse shocks. The setup is described
in detail in Massaglia et al. (\cite{MA05}). The 1D simulation was run on a domain of 
length $L = 4\times 10^{15}\,{\rm cm}$, with a resolution of $1.4\times 10^{11}\,{\rm cm}$,
and the initial velocity perturbation had an amplitude $\Delta v = 30\,{\rm km\,s^{-1}}$.

\begin{figure} [!ht]
  \begin{center}
      \resizebox{\hsize}{!}{\includegraphics{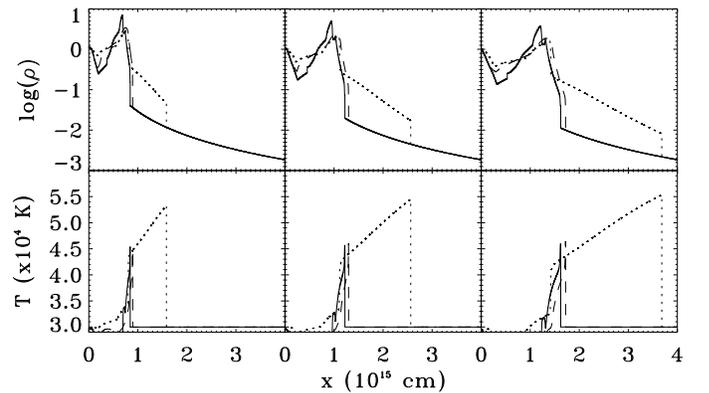}}
    \caption{Density logarithmic profiles (top row) and temperature profiles (bottom row), 
             at three evolutionary stages: adiabatic (dotted line), 
             with SNEq (dashed line), and MINEq (solid line) cooling.}
 \label{fig:evol_1d}
 \end{center}
\end{figure}

In Fig.~\ref{fig:evol_1d}, a comparison is made between the evolution of the formed shock in 
the absence of cooling, with SNEq and the evolution with MINEq. Plots of density and temperature are presented at 
three evolutionary instants in the propagation. 
As it results from the plots in Fig.~\ref{fig:evol_1d}, the shock dynamics are heavily 
influenced by radiative cooling. The shock propagation velocity decreases almost twice in the simulations with 
cooling with respect to the adiabatic ones. The smooth decrease in temperature after the shock front in the
adiabatic simulation is replaced by a much sharper one in the simulations with cooling.

The differences between the density plots obtained with the two cooling models are quantitatively moderate, 
while the shapes are similar. The maximum temperatures attained are very close in the two cooling models. 
Overall, the dynamical differences that appear between the use of the two cooling models are small, 
but the line intensity ratios are very sensitive to density/temperature conditions so the differences 
may result in moderate amplitudes.

The test in equivalent configuration was also performed in a 2D slab. The results were, as expected,
very similar to the ones from the 1D simulations, with somewhat smoother curves due to the lower 
resolution employed. 
It results that for simulations of propagating shocks like the one described it is very important to
include the radiative cooling losses, which heavily influence the dynamics. A simplified treatment of
these losses can be sufficient for studies on the dynamics, while for the computation of emission line 
maps the detailed (MINEq) approach is more suitable. An example is presented in the following section.

%%%%%%%%%%%%%%%%%%%%%%%%%%%%%%%%%%%%%%%%%%%%%%%%%%%%%%%%%%%%%
\subsection{Emission lines}
%
%
%
%
%
%%%%%%%%%%%%%%%%%%%%%%%%%%%%%%%%%%%%%%%%%%%%%%%%%%%%%%%%%%%%%

The computation of emission line ratios from numerical simulations is of great importance
for the field to compare to observations and to discriminate between theoretical
astrophysical models.

In a simple 1D setup, one of the ways to model a YSO jet and to estimate the emission is the following. 
Supposing that the emission comes from shocks inside the jet, the propagation of a shock resulting from 
a velocity fluctuation is simulated in the frame of reference of the jet material. 
The emission in the chosen lines is computed and averaged over a space region corresponding to the resolution
of the observational data (in our example $10^{15} cm$) for each evolutionary step. Then, the resulting averaged
line ratios are plotted at space points corresponding to the transport speed of the jet material, set 
to $150 km\cdot s^{-1}$.

\begin{figure} [!ht]
  \begin{center}
      \resizebox{70mm}{!}{\includegraphics{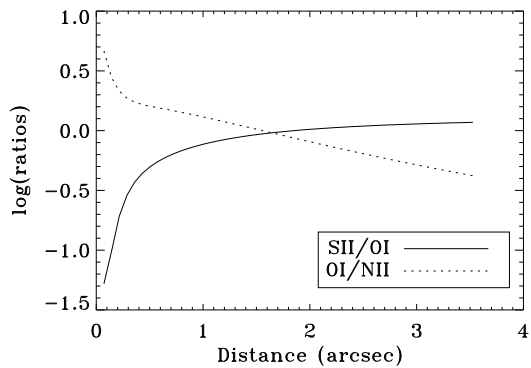}}
    \caption{Line ratios with MINEq cooling, for the propagating shock described in the previous section.}
 \label{fig:lines_1d}
 \end{center}
\end{figure}

The computation was done for the setup presented in Sect. \ref{prop_shocks}. 
The resulting plot, presented in Fig.~\ref{fig:lines_1d}, has also the x axis converted 
in arcseconds (for a distance $D=140pc$) and represents the emission of a jet in the assumption
that the emission comes from internal shocks formed due to initial jet velocity variability.
We present ratios between the forbidden emission lines of \ion{O}{i} $\lambda\lambda 6300\mathrm{\AA}+6364\mathrm{\AA}$, 
\ion{N}{ii} $\lambda\lambda 6548\mathrm{\AA}+6583\mathrm{\AA}$ and \ion{S}{ii} $\lambda\lambda 6716\mathrm{\AA}+6731\mathrm{\AA}$.
Such synthetic emission line ratios can be directly compared with observations of YSO jets.

The main advantage in using MINEq for creating synthetic emission maps is that non-equilibrium ionization 
balance for the elements are provided. The computation of the emission in selected lines is done in post-processing,
with routines distributed together with the code.

%%%%%%%%%%%%%%%%%%%%%%%%%%%%%%%%%%%%%%%%%%%%%%%%%%%%%%%%%%%%%
\subsection{Jet propagation}
%
%
%
%
%
%%%%%%%%%%%%%%%%%%%%%%%%%%%%%%%%%%%%%%%%%%%%%%%%%%%%%%%%%%%%%

Observations of YSO jets that show series of emission knots along their length 
pointed out that simple steady-state models cannot explain their morphology. 
These knots have been interpreted in the literature as due either to the nonlinear
evolution of Kelvin-Helmholtz instabilities set at the jet-ambient interface (Micono et al. \cite{MA00}) 
or to velocity variabilities of the beam (Internal Working Surfaces, see for example Raga et al. \cite{RA02}) 
that, during their propagation, steepen into shocks. 
The latter scenario has been chosen as a possible astrophysical application of the cooling module.

In the present case, we consider a variable jet in 2D cylindrical geometry propagating
into a uniform ambient medium with particle number density $n_{\rm a} = 200 \ \mathrm{cm^{-3}}$
and temperature $T_{\rm a} = 2\,500 \ {\rm K}$.
The beam is injected at the $z=0$, $r<R_j$ ($R_{\rm j} = 2.5 \times 10^{15} \ \mathrm{cm}$) 
boundary with higher density  ($n_{\rm j} = 5 \ n_{\rm a}$) than the background.  
The mean jet velocity is $v_{\rm j}=110 \ \mathrm{km \ s^{-1}}$ with sinusoidal oscillations of amplitude
$\Delta v = 25 \ \mathrm{km \ s^{-1}}$ and a period of $\tau = 50 \ \mathrm{yrs}$

A purely toroidal magnetic field is injected at $z=0$ along with the beam, 
following the simple configuration described in Lind \& al. (\cite{LA89}):
\begin{equation}
\label{eq: B_toroidal}
B_\phi(r) = \left\{\begin{array}{ll}
\DS B_{\rm m} \frac{r}{a} &  \quad{\rm for} \quad 0 \leq r < a        \,,  \\  \noalign{\medskip}
\DS B_{\rm m} \frac{a}{r} &  \quad{\rm for} \quad a\leq r < R_{\rm j} \,,  \\ \noalign{\medskip}
\DS 0                             & \quad{\rm otherwise} \,,
\end{array}\right.
\end{equation}
where $B_{\rm m}$ and $a$ are the magnetization strength and 
radius. 
Demanding pressure equilibrium at the jet inlet, 
$d(p + B_\phi^2)/dr = - B_\phi^2/r$, 
one recovers the pressure profile inside the beam ($r\le R_{\rm j}$),
\begin{equation}
 p(r) = p_0 - B_m^2\min\left(1, \frac{r^2}{a^2}\right)
\end{equation}
where $p_0$ is corresponds to a central temperature 
$T_0 = 10\,000 \ {\rm K}$.
Finally, the magnetization strength, $B_{\rm m}$ is prescribed  from 
the plasma $\beta$ parameter, defined in terms of the averaged beam 
pressure:
\begin{equation}
 \beta \equiv \frac{2}{B_{\rm m}^2}\, 
         \frac{\int_0^{R_j} p(r) r\, dr}{\int_0^{R_j} rdr}
    = \frac{a^2}{R_{\rm j}^2} + 2\frac{p_0}{B_m^2} - 2
\end{equation}
from which one can easily recover $B_{\rm m}$.
For the present computation, we set $a = 0.6 R_{\rm j}$
and $\beta = 1$.
This choice of parameters is similar to the ones found by Masciardi \& Raga 
(\cite{MR01}) in their attempt to model the curved HH 505 jet.

Numerical integration is carried out with the PPM method 
and the HLLC Riemann solver of Li (\cite{Li05}). We use
$30$ zones on the jet radius and the domain extends, in beam radii, from 
$0$ to $10$ in the radial direction and from $0$ to $60$ in the
longitudinal direction.
Free outflow is assumed across the outer boundaries, whereas reflecting 
boundary conditions hold at the axis ($R=0$) and outside the jet
inlet ($z = 0, r > R_{\rm j}$).
A smoothing function is introduced for all variables at the transition between the jet material 
and the external medium to avoid the formation of an unphysical high temperature low-density 
layer around the jet.

\begin{figure*} [!ht]
 \begin{center}
 \resizebox{165mm}{!}{\includegraphics{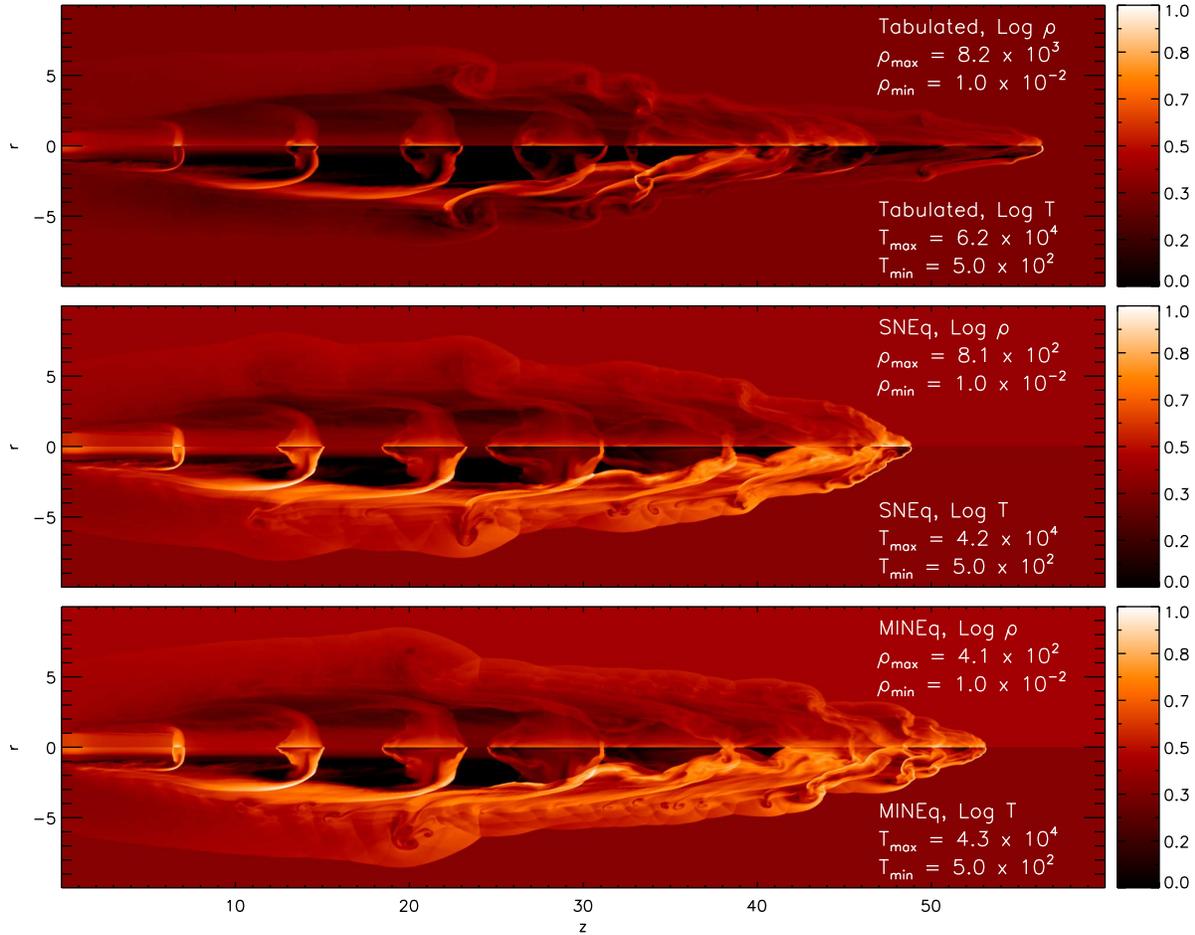}}
 \caption{Tabulated Cloudy (top panel), radiative with SNEq cooling (middle panel) and MINEq cooling 
          (bottom panel) for the jet simulation. In each panel, the upper and lower halves show
          density and temperature in log scale, respectively. The different shades are normalized
          between $\rho_{\min}$ and $\rho_{\max}$ (for density), $T_{\min}$ and $T_{\max}$ 
          (for temperature).}
 \label{fig:2djet}
 \end{center}
\end{figure*}

We perform a set of three simulations, by adopting i) a tabulated cooling function, ii) simplified
treatment of radiative cooling losses (SNEq), and iii) the detailed cooling treatment.
The tabulated cooling function simply consists in adding a source term to
the energy equation, given by the Cloudy Z=1 cooling curve (presented in Sect. \ref{effcool}) as a
tabulated function of temperature. This cooling implementation does not follow the ionization balance of 
any element, but as a standard procedure for this kind of approach, the effective cooling function is 
multiplied by the particle density squared to obtain energy losses per unit volume.
Results at $t\approx 500 \mathrm{yrs}$ are shown in Fig. (\ref{fig:2djet}).

The pulsed initial jet velocity produces, as expected, a number of intermediate 
shocks propagating along the jet with typical post-shock temperatures
in the range $15\,000 \div 25\,000$ K. 
The morphology is similar between the SNEq and MINEq runs, since
at these temperatures the two cooling losses are comparable.
Larger deviations are observed close to the head of the jet, where
temperatures are higher and the two cooling functions exhibit larger
differences.
Overall, while SNEq and MINEq give similar results,
radiated losses are higher for the tabulated Cloudy cooling, as can be inferred by the 
reduced lateral expansion of the cocoon (this can be expected considering the 
effective cooling curves in Fig. \ref{fig:total_cool}).

Nevertheless, temperatures at the jet head are highest for the tabulated cooling case.
In order to understand this apparently unexpected result, we 
have performed a systematic comparison between the Tabulated and SNEq cooling functions 
by means of supplementary simulations (not shown here).
Our results demonstrate that at high temperatures ($\gtrsim 4\cdot 10^4$ K) and low 
ionization, the SNEq line emission becomes larger than the equilibrium Cloudy one.
In this case, it is crucial that the SNEq line emissions depend on the electron number density 
and that this density is dynamically computed from the non-equilibrium ionization of H. 
In the tabulated case, an equilibrium ionization balance is implicitly assumed.
This confirms that the maximum temperatures can be an indication
of the maximum cooling losses attained locally, but not on the overall cooling losses.
It must also be noticed that the resolution of these simulations is still low to resolve the
post-shock zone at the jet head, so the maximum temperatures observed may be subject to 
large uncertainties.

We conclude that even a simple cooling like SNEq, evolving only the hydrogen fraction, is a much 
better approximation than using tabulated cooling losses.

\begin{figure*} [!ht]
 \begin{center}
 \resizebox{165mm}{!}{\includegraphics{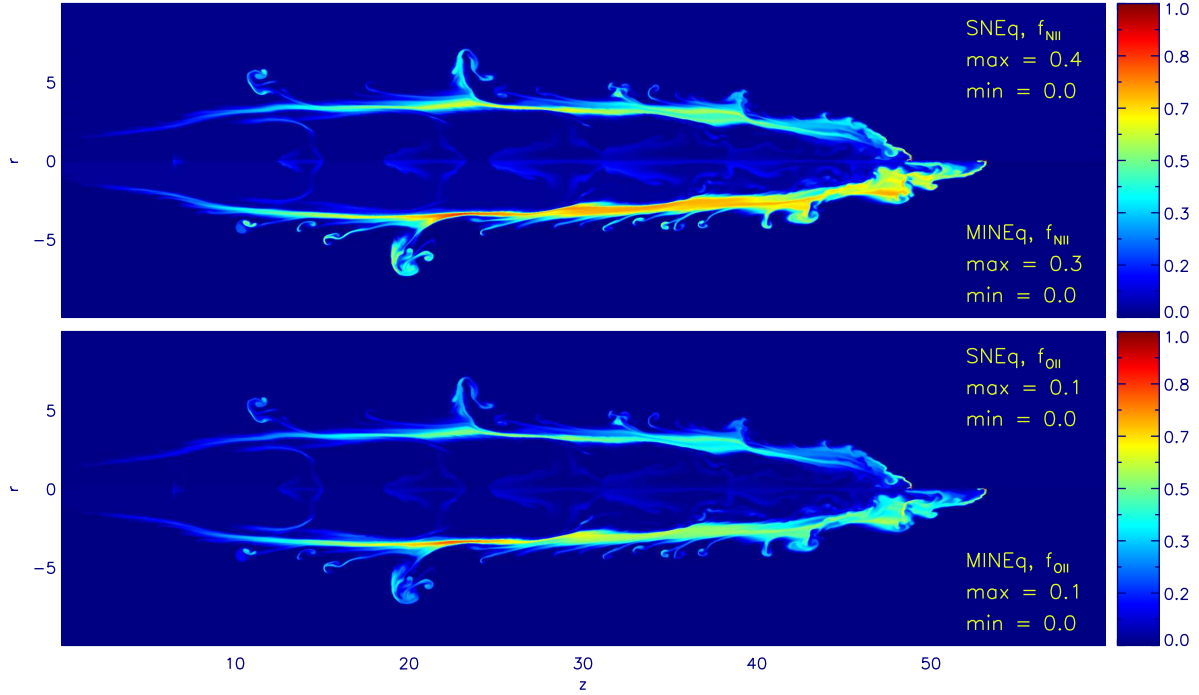}}
 \caption{Fractions of \ion{N}{ii} (top panel) and \ion{O}{ii} 
          (bottom panel) for the jet simulation. In each panel, the upper and lower halves show
          the results with SNEq and MINEq, respectively. The different shades are normalized
          between the minimum and maximum values of the fractions.}
 \label{fig:2djet_NO}
 \end{center}
\end{figure*}

The ionization fractions computation is very important when it comes to producing synthetic emission
maps in various emission lines to be compared with observations. In Fig. (\ref{fig:2djet_NO}),
the fractions of \ion{N}{ii} and \ion{O}{ii} are presented, dynamically computed by MINEq and
alternatively computed from SNEq considering them fixed by the hydrogen ionization through charge-transfer 
(see Osterbrock \& Ferland \cite{OF05}). 
The differences are moderate and can result in variations of $20-30\%$ in the emission
lines computation.

The steep gradients and transition regions forming immediately
behind the shock front are crucial in determining the emission properties, 
e.g. line intensity ratios. For this reason they need to be accurately resolved 
(Massaglia et al. \cite{MA05}).
However, at the resolution employed here ($300\times 1800$), only the general physical
evolution can be captured. Considerably higher resolution is required 
and this can be efficiently achieved only through adaptive mesh refinement 
simulations. This issue will be the subject of forthcoming works.

%%%%%%%%%%%%%%%%%%%%%%%%%%%%%%%%%%%%%%%%%%%%%%%%%%%%%%%%%%%%%
\section{Conclusions}
%
%
%
%
%
%%%%%%%%%%%%%%%%%%%%%%%%%%%%%%%%%%%%%%%%%%%%%%%%%%%%%%%%%%%%%

After a series of tests and validation, the detailed treatment of radiative losses MINEq is now implemented
in the MHD code PLUTO. The choice of the integration technique and the optimizations are, as far as we
know, unique at the time of this writing and provide a high degree of accuracy in ion species abundances 
and radiative losses computation. Both theoretical and technical aspects of the current implementation,
as well as testing process and applications were discussed in the previous sections.

A major advantage obtained by using MINEq, compared to previously employed SNEq, 
is that the line emission can be computed in conditions of non-equilibrium ionization for all species,
more likely to be encountered in situations of rapid changes, as it is the case of shock waves.

As shown by the tests presented, the choice of the cooling model
between MINEq and SNEq, has an increasing effect on the structure 
formation whenever temperatures exceed $25\,000$ K, and has an important
influence for the ionization fractions that reflects in the emission line computations.
For a preliminary dynamical study, a tabulated cooling function that does not integrate any ion specie
can be employed, but the relatively low computational cost of a cooling that evolves the 
hydrogen ionization in a time-dependent fashion makes the latter advisable in most cases.
Whenever the purpose is the computation of emission line ratios, employing a detailed cooling
like MINEq produces more reliable results as the ionization fractions are followed in 
non-equilibrium conditions, twhich are likely to be encountered in the real astrophysical objects.

An important feature of the cooling model implementation is that it is upgradeable, more ion species and 
other processes can be added (increasing also the temperature range of application). Also, it can be 
used as starting point for the integration of atomic chemistry and cooling processes to other MHD codes.

The newly-developed cooling function provides a powerful tool for investigating the stellar jets and
gaseous nebulae. It is, in the current configuration, suited for the study of radiative shocks in 
stellar jets. 
High resolution, adaptive numerical simulations to predict line emission in 
YSO jets will be subject of forthcoming works.

%%%%%%%%%%%%%%%%%%%%%%%%%%%%%%%%%%%%%%%%%%%%%%%%%%%%%%%%%%%%%
\appendix
%
%
%
%
%
%%%%%%%%%%%%%%%%%%%%%%%%%%%%%%%%%%%%%%%%%%%%%%%%%%%%%%%%%%%%%

%%%%%%%%%%%%%%%%%%%%%%%%%%%%%%%%%%%%%%%%%%%%%%%%%%%%%%%%%%%%%
\section{Ionization/recombination processes}
\label{append_ionrec}
%
%
%
%
%
%%%%%%%%%%%%%%%%%%%%%%%%%%%%%%%%%%%%%%%%%%%%%%%%%%%%%%%%%%%%%

The following processes are taken into consideration: collisional ionization, radiative, and dielectronic 
recombination, charge-transfer with H and He. These processes enter the ionization/recombination coefficients
defined and used in Sect. \ref{cfdesc}.

\subsection{Collisional ionization}

We use the Voronov (\cite{Vo97}) data to estimate the collisional ionization rates with the analytical formula:
\begin{equation}
\zeta ^{\mathrm{coll}} = A \cdot \frac{ 1 + P \cdot U ^{1/2} }{ X + U } \cdot U^{\mathrm{K}} \cdot e^{\mathrm{-U}},
\end{equation}
where $ U = \Delta E / T $ and $A$, $P$, $\Delta E$, $X$, and $K$ parameters are listed in Table 1
of the cited paper. 
$T$ and $\Delta E$ are expressed in eV, and $\zeta^{\mathrm{coll}}$ in $\mathrm{cm^3 s^{-1}}$.

The actual number of ionizations in unit time and unit volume will be:
\begin{equation}
\frac{dN}{dt} = N_i \cdot N_{\mathrm{el}} \cdot \zeta^{\mathrm{coll}}
\end{equation}
where $N_{i}$ is the total number density of atoms in the lower-ionization state, 
and $N_{\mathrm{el}}$ the electron number density.

\subsection{Radiative recombination}

The total radiative recombination rates are taken from P\`{e}quignot et al. (\cite{PA91}). The total recombination rate is
fitted with the analytical formula:
\begin{equation}
\alpha^{\mathrm{RR}} = 10^{-13} z \frac{a t^b}{1 + c t^d},
\end{equation}
where $z$ is the ionic charge and $t = 10^{-4} T(K) / z^2$. The four parameters $a$, $b$, $c$, and $d$ are given in Table 1 
in the cited paper.

The resulting $\alpha^{\mathrm{RR}}$ is expressed in units of $\mathrm{cm^3 s^{-1}}$.

\subsection{Dielectronic recombination}

The dielectronic recombination process proceeds as
\begin{equation}
A_p^{\mathrm{+m+1}} + e^- \rightarrow A_a^{\mathrm{+m}} \rightarrow A_b^{\mathrm{+m}} + h \nu ,
\end{equation}
where $p$ stands for a state of the $m+1$ times ionized element $A$, and $a$ and $b$ represent an auto-ionizing and a true
bound state of the next ionization stage.

From Nussbaumer \& Storey (\cite{NS83}), the dielectronic recombination rates are fitted by the analytical formula:
\begin{equation}
\alpha^{\mathrm{DR}} = 10^{-12} \left( \frac{a}{t} + b + ct + c t^2 \right) t^{-3/2} \exp{\left(\frac{-f}{t}\right)} ,
\end{equation}
where $t = T(K) / 10^4K$ and $\alpha^{\mathrm{DR}}$ is expressed in $\mathrm{cm^3 s^{-1}}$. The coefficients are given in a 
table from the cited paper.
We used the data from Nussbaumer \& Storey (\cite{NS83}) for the C, N, and O ions.

\subsection{Total electron - ion recombination}

For the He, Ne, and S ions, data from Kato \& Asano (\cite{KA99}) was used for the total recombination coefficient 
(radiative + dielectronic). These are tabulated values that we interpolate in our temperature range. 

\subsection{Charge transfer with H}

The charge transfer (exchange) reactions with H are reactions of the form:
\begin{equation}
A^{\mathrm{+n}} + H \rightleftarrows A^{\mathrm{+(n-1)}} + H^+ + \delta E
\end{equation}
The direct reaction is called charge-transfer recombination and the inverse charge-transfer ionization ($\zeta^{\rm HII}$).

We took the data for charge transfer with H from Kingdon \& Ferland (\cite{KF96}). The recombination/ionization rate is fitted by
the analytical formula
\begin{equation}
\alpha^{\mathrm{HI}} , \zeta^{\rm HII} = a t_4^b [ 1 + c e^{d t_4} ],
\end{equation}
where $t_4 = T(\mathrm{K}) / 10^4 \mathrm{K}$ and the parameters $a$, $b$, $c$, and $d$ are listed in Tables 1 and 3 
from the cited paper.

\subsection{Charge transfer with He}

The charge transfer (exchange) reactions with He are reactions of the form:
\begin{equation}
A^{\mathrm{+n}} + He \rightarrow A^{\mathrm{+(n-1)}} + He^+ \,.
\end{equation}

We took the data for charge transfer with He from Wang et al. (\cite{WA01}) and references herein. The recombination 
rate is fitted by the analytical formula
\begin{equation}
\alpha^{\mathrm{HeI}} = a t_4^b [ 1 + c \exp{(d t_4)} ],
\end{equation}
where $t_4 = T / 10^4 \mathrm{K}$ and the parameters $a$, $b$, $c$, and $d$ are listed in tables available on-line.

%%%%%%%%%%%%%%%%%%%%%%%%%%%%%%%%%%%%%%%%%%%%%%%%%%%%%%%%%%%%%
\section{Jacobian Matrix}
\label{append_impl}
%
%
%
%
%
%%%%%%%%%%%%%%%%%%%%%%%%%%%%%%%%%%%%%%%%%%%%%%%%%%%%%%%%%%%%%

The solution of implicitly linearized equations such as Eq. 
(\ref{eq:semi-impl}) or the Rosenbrock scheme requires the
expression of the Jacobian matrix of the system of equations
given by (\ref{eq:source_ode}).
Using the definition,
\begin{equation}
 \tens{J} \equiv \pd{\vec{f}(\vec{y})}{\vec{y}} = \left( \begin{array}{cc}
\DS  \pd{\vec{\dot{X}}}{\vec{X}} & \DS\pd{\vec{\dot{X}}}{p} \\  \noalign{\medskip}
\DS  \pd{\dot{p}}{\vec{X}} & \DS\pd{\dot{p}}{p}
\DS  \end{array} \right)
\end{equation}
where $\vec{y} = \{p, \vec{X}\}$ and $\vec{f}(\vec{y}) \equiv \dot{\vec{y}}$.
Partial derivatives are computed using combined analytical and 
numerical differentiation. 
We note in the first place that, for practical reasons, the right-hand 
side $\vec{f}(p,\vec{X})$ is better expressed in terms of temperature 
and ionization fractions, that is
\begin{equation}
  \vec{f}\Big(p, \vec{X}\Big) \equiv
  \vec{g}\Big(T, \vec{X}\Big)  \,,
\end{equation}
where $T$, $p$, and $\vec{X}$ are related through
\begin{equation}\label{eq:temperature}
 T = \frac{p}{\rho}\frac{m_u\mu(\vec{X})}{k_B} \,.
\end{equation}
This allows us to compute partial derivatives with respect
to the ion fractions using the chain rule, 
\begin{equation}\label{eq:partial_X}
  \left.\pd{\vec{f}}{X_{\xi,m}}\right|_p = 
  \left.\pd{\vec{g}}{X_{\xi,m}}\right|_T + 
  \left.\pd{\vec{g}}{T}\right|_{\vec{X}}\pd{T}{X_{\xi,m}} \,,
\end{equation}
where, using Eq. (\ref{eq:temperature}) and the definitions of the mean
molecular weight, Eq. (\ref{eq:mu}), we can express the second
term on the right as
\begin{equation}\label{eq:2ndterm}
  \left.\pd{\vec{g}}{T}\right|_{\vec{X}}\pd{T}{X_{\xi,m}} =
  \left.\pd{\vec{f}}{p}\right|_{\vec{X}}\frac{p}{\mu}
\left[\frac{1}{\mu_D}\pd{\mu_N}{X_{\xi,m}} - \frac{\mu}{\mu_D}\pd{\mu_D}{X_{\xi,m}}\right]\,.
\end{equation}
where the term is square brackets is simply $\partial\mu(\vec{X})/\partial X_{\xi,m}$ 
whereas $\mu_N$ and $\mu_D$ are, respectively, the numerator and the 
denominator of the mean molecular weight.

The explicit dependence on $T$ and $\vec{X}$ in our 
ionization network, Eq. (\ref{eq:ion_bal}), is made clear 
by rearranging terms as 
\begin{equation}\begin{split}
 \dot{X}_{\kappa,i} =   L_{\kappa,i}\left(T, \vec{X}\right) X_{\kappa,i-1}  
             & - C_{\kappa,i}\left(T, \vec{X}\right) X_{\kappa,i} + \\
             & + R_{\kappa,i}\left(T, \vec{X}\right) X_{\kappa,i+1} \,,
\end{split}\end{equation}
for each element's ions.
The coefficients $L_{\kappa,i}, C_{\kappa,i}$, and $R_{\kappa,i}$ are expressed by 
sums of functions depending on either $T$ or $\vec{X}$:
\begin{eqnarray}
 L_{\kappa,i} & = & L^a_{\kappa,i}N_{\rm el} + L^b_{\kappa,i}X_{\rm HI} + L^c_{\kappa,i} X_{\rm HeI} + L^d_{\kappa,i} \\ 
  \noalign{\medskip}
 C_{\kappa,i} & = & C^a_{\kappa,i}N_{\rm el} + C^b_{\kappa,i}X_{\rm HI} + C^c_{\kappa,i} X_{\rm HeI} + C^d_{\kappa,i} \\
  \noalign{\medskip}
 R_{\kappa,i} & = & R^a_{\kappa,i}N_{\rm el} + R^b_{\kappa,i}X_{\rm HI} + R^c_{\kappa,i} X_{\rm HeI} + R^d_{\kappa,i} 
\end{eqnarray}
where the $L^{\cdots}_{\kappa,i}$'s, $C^{\cdots}_{\kappa,i}$'s, and $R^{\cdots}_{\kappa,i}$'s depend 
on $T$ only whereas $N_{\rm el}$, given by Eq. (\ref{eq:nel}), depends
on $\vec{X}$ only.
Focusing on the Jacobian sub-matrix $\partial\dot{\vec{X}}/\partial\vec{X}$,
we evaluate the first term in Eq. (\ref{eq:partial_X}) as 
\begin{equation}\begin{split}
  \left.\pd{\dot{X}_{\kappa,i}}{X_{\xi,m}}\right|_T =  
              L_{\kappa,i}\delta_{i-1,m}\delta_{\kappa,\xi} & + \pd{L_{\kappa,i}}{X_{\xi,m}}X_{\kappa,i-1} - \\
	      - C_{\kappa,i}\delta_{i,m}\delta_{\kappa,\xi} & - \pd{C_{\kappa,i}}{X_{\xi,m}}X_{\kappa,i} + \\
              + R_{\kappa,i}\delta_{i+1,m}\delta_{\kappa,\xi} + \pd{R_{\kappa,i}}{X_{\xi,m}X_{\kappa,i+1}} 
\end{split}\end{equation}
where $\delta_{i,m}$ is the Kronecker delta symbol and
\begin{eqnarray}
  \pd{L_{\kappa,i}}{X_{\xi,m}} & = & L^a_{\kappa,i}(T) N\gamma_mA_{\xi} + L^b_{\kappa,i}(T) \delta_{HI,\xi m} + \nonumber \\
		     & & + L^c_{\kappa,i}(T)\delta_{HeI,\xi m}\\
 \noalign{\medskip}
  \pd{C_{\kappa,i}}{X_{\xi,m}} &= &C^a_{\kappa,i}(T) N\gamma_mA_{\xi} + C^b_{\kappa,i}(T) \delta_{HI,\xi m} +  \nonumber \\
                     & & + C^c_{\kappa,i}(T)\delta_{HeI,\xi m} \\
 \noalign{\medskip}
 \pd{R_{\kappa,i}}{X_{\xi,m}} & = & R^a_{\kappa,i}(T) N\gamma_mA_{\xi} + R^b_{\kappa,i}(T) \delta_{HI,\xi m} +  \nonumber \\
                     & & + R^c_{\kappa,i}(T)\delta_{HeI,\xi m} 
\end{eqnarray}
In the previous equations we made use of the fact that
 $\partial N_{\rm el}/\partial X_{\kappa,i} = N \left( \gamma_i - 1 \right) B_{\kappa}$.

The last row of the Jacobian involves derivatives of the
cooling function with respect to $\vec{X}$: 
\begin{equation}
  \pd{\dot{p}}{X_{\xi,m}} = - N_{\rm at}\gamma_m B_{\xi}\frac{\Lambda}{N_{\rm el}}  
                      - N_{\rm at}N_{\rm el} \pd{\Lambda}{X_{\xi,m}} - \pd{L_{FF}}{X_{\xi,m}}
                      - \pd{L_{I-R}}{X_{\xi,m}}
\end{equation}
where 
\begin{equation}
 \pd{\Lambda}{X_{\xi,m}} = {\cal L}_{\xi,m}B_{\xi} + 
 N\gamma_mB_{\xi}\sum_{\kappa,i} X_{\kappa,i} \pd{{\cal L}_{\kappa,i}}{X_{\xi,m}}B_{\beta} \,,
\end{equation}
$\partial{\cal L}_{\kappa,i}/\partial{X_{\xi,m}}$ is found numerically 
and the remaining terms are found by straightforward differentiation of
the energy losses due to ionization-recombination and bremsstrahlung.

Finally, partial derivatives with respect to pressure needed in 
Eq. (\ref{eq:2ndterm}) and in the last column of $\tens{J}$ 
are computed numerically using a centered approximation:
\begin{equation}
 \pd{f}{p} \approx \frac{f\left(p(1+\epsilon),\vec{X}\right) - f\left(p(1-\epsilon),\vec{X}\right)}
                        { \epsilon p}
\end{equation}
where $\epsilon$ is a small parameter (typically $\epsilon = 10^{-4}$).

\begin{acknowledgements}
This work has been supported by the EU contract MRTN-CT-2004-005592 within the Marie Curie RTN JETSET.\\
The Cloudy curves were created by Michal Rozyczka and Guillermo Tenorio-Tagle and later 
updated by Tomek Plewa with the help of version 90.01. OT wishes to thank Dr. Tomek Plewa for 
comments on the Cloudy results. We thank the referee for valuable comments and observations
useful in improving the paper.
\end{acknowledgements}


\begin{thebibliography}{}

\bibitem[2004]{Ba04}
  Bacciotti, F. 2004, 
  \apjs, 293, 37 
  
\bibitem[1993]{BL93}
  Blondin, J.M., Lufkin, E.A. 1993,
  \apjs, 88, 589
  
\bibitem[1992]{BP92}
  Blum, R.D., Pradhan, A.K. 1992
  \apjs, 80, 425
  
\bibitem[1998]{BL98}
  Berger, M.J., \& LeVeque, R.J. 1998,
  SIAM Journal on Numerical Analysis, 35, 6, pp 2298-2316
  
\bibitem[1972]{DM72}
  Dalgarno, A., \& McCray, R.A. 1972,
  \araa, 10, 375

\bibitem[1998]{EOO98}
   Ekeland, K., Owren, B. and  {\O}ines E., 1998, 
  {ACM} Transactions on Mathematical Software, 24, 368

\bibitem[1998]{FA98}
  Ferland, G.J., Korista, K.T., Verner, D.A., et al. 1998,
  PASP, 110, 761

\bibitem[1987]{GA87}
  Giovanardi, C., Natta, A., Palla, F. 1987,
  \aaps, 70, 269

\bibitem[1999]{KA99}
  Kato T., Asano E. 1999, 
  National Institute for Fusion Science - Japan, NIFS-DATA-54

\bibitem[1996]{KF96}
  Kingdon J.B., Ferland G.J. 1996,
  \apjs, 106, 205

\bibitem[2000]{LA00}
  Lavalley-Fouquet, C., Cabrit, S., \& Dougados, C. 2000,
  \aap, 356, L41

\bibitem[1984]{LW84}
  Leahy, J.P., \& Williams, A.G.
  1984, \mnras, 210, 929

\bibitem[2005]{Li05}
  Li, Shengtai 2005,
  Journal of Computational Physics, 203, 344 
  
\bibitem[1989]{LA89}
  Lind, H., Payne, D., Meier, D., \& Blandford, R. 1989,
  \apj, 344, 89 
  
\bibitem[1997]{MS97}
  Marten, H., Szczerba, R. 1997,
  \aap, 325, 1132
  
\bibitem[2001]{MR01}
  Masciadri, E., \& Raga, A.C.,  2001,
  \aj, 121, 408

\bibitem[2005]{MA05}
  Massaglia, S., Mignone, A., \& Bodo, G. 2005,
  \aap, 442, 549

\bibitem[2000]{MA00}
  Micono, M., Bodo, G., Massaglia, S., et al. 2000,
  \aap, 360, 795
  
\bibitem[2007]{MA07}
  Mignone, A., Massaglia, S., Bodo, G., et al. 2007,
  \apjs, 170, 228
   
\bibitem[1983]{NS83}
  Nussbaumer H., Storey P.J. 1983,
  \aap, 125, 75

\bibitem[2005]{OF05}
  Osterbrock D.E., \& Ferland G.J. 2005,
  Astrophysics of Gaseous Nebulae and Active Galactic Nuclei,
  University Science Books

\bibitem[1991]{PA91}
  P\`{e}quignot D., Petitjean, P.; Boisson, C. 1991,
  \aap, 251, 680
  
\bibitem[2005]{PA05}
  Poludnenko, A., Varni\`ere, P., Cunningham, A., Frank, A., Mitran, S. 2005,
  Lecture Notes in Computational Science and Engineering 41, pp 331-340,
  Springer

\bibitem[1999]{PZ99}
  Pradhan, A.K., \& Zhang, H.L. 1999,
  Landolt-Boernstein Volume 17: Photon and Electron Interactions with Atoms, Molecules, Ions ,
  Springer-Verlag, I.17.B, pp 1-102

\bibitem[1992]{NR}
  Press W.H., Flannery B.P., Teukolsky S.A., Vetterling W.T. 1992
  Numerical Recipes in C, Cambridge University Press

\bibitem[1997]{RM97}
  Raga A.C., Mellema G., Lundqvist P. 1997, 
  \apjs, 109, 517

\bibitem[2000]{RA00}
  Raga, A.C., Navarro-Gonz\`alez, R., Villagr\`an-Muniz, M. 2000,
  \rmxaa, 36, 67

\bibitem[2002]{RA02}
  Raga, A.C., Vel\`azquez, P.F.,  Cant\'{o}, J., Masciadri, E. 2000,
  \aap, 395, 647

\bibitem[1992]{Ra92}
  Raymond J.C. 1992, 
  Private communication  
  
\bibitem[1997]{RA97}
  Rossi, P., Bodo, G., Massaglia, S., Ferrari, A. 1997,
  \aap, 321, 672

\bibitem[1993]{ST93}
  Schmutzler, T., \& Tscharnutter, W.M. 1993,
  \aap, 273, 318
  
\bibitem[1994]{SA94}
  Stafford, R.P., Bell, K.L., Hibbert, A., Wijesundera, W.P. 1994,
  \mnras, 268, 816

\bibitem[1993]{SD93}
  Sutherland, R.S., Dopita, M.A. 1993, 
  \apjs, 88, 253
  
\bibitem[2003]{SA03}
  Sutherland, R.S., Bicknell, G.V., Dopita, M.A. 2003,
  \apj, 591, 238

\bibitem[1997]{Vo97}
  Voronov G.S. 1997,
  Atomic Data and Nuclear Data Tables 65, 1-35

\bibitem[2001]{WA01}
  Wang et al. 2001,
  ORNL/UGA Charge Transfer Database for Astrophysics,
  http://cfadc.phy.ornl.gov/astro/ps/data/

 
\end{thebibliography}
\end{document}